\documentclass{elsart}
\input{psfig.sty}

\usepackage{natbib}

\usepackage{amssymb}

\begin{document}

\begin{frontmatter}



\title{Why the fundamental plane of black hole activity is not simply
  a distance driven artifact}


\author[am]{A. Merloni}
\author[ek]{E. K\"ording}
\author[sh]{S. Heinz\thanksref{ch}}
\author[sm]{S. Markoff}
\author[tdm]{T. Di Matteo}
\author[hf]{H. Falcke}
\address[am]{Max-Planck-Institut f\"ur Astrophysik,
  Karl-Schwarzschild-Strasse 1, D-85741, Garching, Germany, e-mail:
  {\tt am@mpa-garching.mpg.de}}
\address[ek]{School of Physics \& Astronomy, University of Southampton,
Southampton, SO17 1BJ UK}
\address[sh]{Kavli Institute for Astrophysics and Space Research, MIT, 77
  Mass. Ave., Cambridge, MA 02139}
\address[sm]{Astronomical Institute "Anton Pannekoek", University of Amsterdam, Kruislaan 403, 1098 SJ, Amsterdam, the, Netherlands}
\address[tdm]{Carnegie Mellon University, Department of Physics, 5000 Forbes
  Ave., Pittsburgh, PA 15213}
\address[hf]{Max-Planck-Institut f\"ur Radioastronomie, Auf dem H\"ugel 69,
53121 Bonn, Germany; Radio Observatory, ASTRON, Dwingeloo, PO Box 2,
7990 AA Dwingeloo, The Netherlands; Dept. of Astronomy, Radboud
Universiteit Nijmegen, Postbus 9010, 6500 GL Nijmegen, The
Netherlands}

\thanks[ch]{Chandra Fellow}

\begin{abstract}
The fundamental plane of black hole activity is a non-linear
correlation among radio core luminosity, X-ray luminosity and mass of
all accreting black holes, both of stellar mass and supermassive,
found by Merloni, Heinz and Di Matteo (2003) and, independently, by
Falcke, K\"ording and Markoff (2004). Here we further examine a number
of statistical issues related to this correlation. In particular, we
discuss the issue of sample selection and quantify the bias introduced
by the effect of distance in two of the correlated quantities. We
demonstrate that the fundamental plane relation cannot be a distance
artifact, and that its non-linearity must represent an intrinsic
characteristic of accreting black holes. We also discuss possible
future observational strategies to improve our understanding of this
correlation.
\end{abstract}


\end{frontmatter}

\section{Introduction}
\label{sec:introduction}
The search for statistical associations between the X-rays and radio
core emission in Quasars and AGN is about as old as X-ray astronomy
itself. Very early on, a number of statistical issues related to the
search of correlations between radio and X-ray luminosities in
actively accreting black holes was already under discussion. In fact,
the questions that arose in this discussion stimulated the
formulation and the wider recognition of a set of statistical methods
specifically targeted to astrophysical problems
(see e.g. Feigelson and Berg 1983; Kembhavi, Feigelson and Singh 1986;
for a comprehensive discussion of statistical methods and problems in
astrophysics, see Babu and Feigelson 1996 and references therein).

In particular, the fundamental question was raised (see e.g.~Elvis et
al.~1981; Feigelson and Berg 1983) of whether correlations are more
accurately measured by comparing observed flux densities or intrinsic
luminosities, as it is obvious that in flux limited samples spanning
large ranges in redshift (i.e.~distance) spurious correlations can be
inferred in luminosity-luminosity plots if only detected points are
considered.  On the other hand, as clearly discussed in Feigelson and
Berg (1983) and in Kembhavi et al.~(1986), flux-flux correlations can
themselves lead to spurious results, whenever there exists any
non-linear intrinsic correlation between luminosities. The most
statistically sound way to deal with the aforementioned biases has
been formalized in terms of partial correlation analysis capable to
handle censored data (upper limits), as discussed in Akritas and
Siebert (1996). With such a method not only a correlation coefficient
can be calculated for any luminosity-luminosity relationship in flux
limited samples, but also a significance level can be assigned to it.

\subsection{The fundamental plane of black hole activity}
\label{sec:fp}
Black holes as mathematical entities are extremely simple, being fully
described by just three quantities: mass, spin and charge.  For
astrophysical black holes, necessarily uncharged, little is known so
far about their spin distribution. However, it is well established
observationally that black holes do span a wide range in masses, from
the $\sim 10 M_{\odot}$ ones in X-ray Binaries (XRB) to the
supermassive ($\sim 10^6 - 10^9 M_{\odot}$) ones in the nuclei of
nearby galaxies and in Active Galactic Nuclei (AGN). In Merloni, Heinz
and Di Matteo (2003; MHD03), we posed the following question: is the
mathematical simplicity of black holes also manifest in their
observational properties? More specifically: which observed black hole
characteristics do scale with mass?

To answer such a question, we searched for a common relation between
X-ray luminosity, radio core luminosity and black hole masses among
X-ray binaries and AGN.  This necessarily imposes a set of
complications for any statistical analysis. These are essentially
twofold. On the one hand, as already pointed out in the original
papers on the subject (MHD03; Falcke, K\"ording and Markoff 2004,
hereafter FKM04),
there is a vastly differing distance scale between the two populations
that should at some level induce spurious (linear) correlations
between the observed luminosities even for a {\em completely} random
distribution of fluxes.

Moreover, the inclusion of black hole mass in the analysis
imposes some complex selection criterion on any sample: mass is
estimated in a number of different ways, implying different levels and types of
uncertainties linked to the specific observational strategy. It
is therefore virtually impossible, at least with the current data, to
estimate the degree of incompleteness of any black hole mass sample.

\begin{figure}
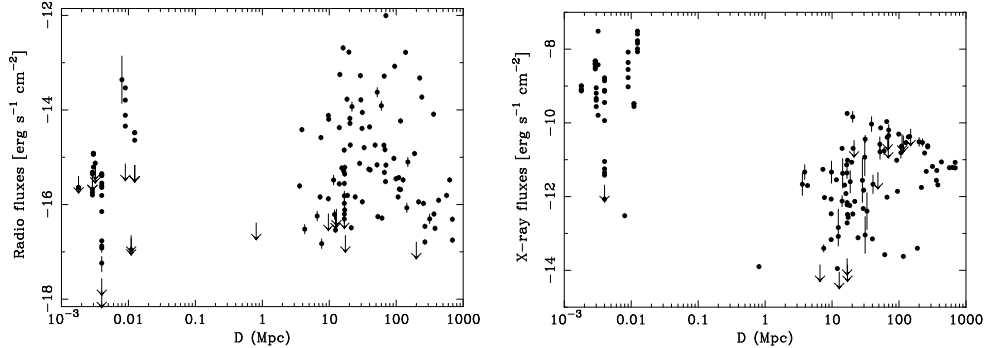

\begin{tabular}{cc}
\psfig{figure=f1a.ps,angle=270,width=0.45\textwidth}&
\psfig{figure=f1b.ps,angle=270,width=0.45\textwidth}\\
\end{tabular}
\caption{Radio (left) and X-ray (right) fluxes for the sources in the
original sample of MHD03 vs.~distance (in Mpc). Upper limits are
marked by green arrows.}
\label{fig:sample}
\end{figure}

For the specific example we are interested in, a relationship is
posited between the radio core luminosity (at 5GHz) $L_{\rm R}$ of a
black hole, its X-ray luminosity $L_{\rm X}$, and its mass $M$.
$L_{\rm R}$ and $L_{\rm X}$ are derived quantities, each carrying, in
addition to the respective flux, a factor of $D^2$, where $D$ is the
luminosity distance to the source.

As discussed in the introduction, statistical tools exist to test
whether a correlation is, in fact, an artifact of distance, or whether
it reflects an underlying luminosity-luminosity relation, even in
flux limited samples \citep{feigelson:83}.  In MHD03 (section 3) a
partial correlation analysis was performed, including all upper limits
in the sample using the algorithm for performing Kendall's $\tau$ test
in the presence of censored data proposed by \cite{akritas:96}. 
Such an analysis showed unequivocally that, even after the large range of
distances in the sample was taken into account, the radio core
luminosity was correlated with both X-ray luminosity and mass, both
for the entire sample (XRB plus AGN) and for the sample of
supermassive black holes only\footnote{The partial correlation
  analysis carried on in MHD03
further demonstrated that the radio core luminosity is correlated with
black hole mass after the common dependence on X-ray luminosity is
taken into account, and vice versa, thus not only justifying, but
statistically {\em mandating} the multivariate linear regression,
rather than just a bivariate one.}.

Motivated by the findings of the partial correlation analysis,
MHD03 performed a linear regression fit to the data and found them to
be well described by the following expression:
\begin{equation}
\label{eq:fp}
Log L_{\rm R} = 0.6 Log L_{\rm X} + 0.78 Log M + 7.33,
\end{equation}
with a substantial residual scatter ($\sigma \simeq 0.88$).  A very
similar result was obtained independently, from a different but
largely congruent sample of sources, at essentially the same time by
FKM04.

In the following, we review some of the original arguments presented
in MHD03 that address the following question: is the multivariate
correlation of Eq.(\ref{eq:fp}) a spurious result due to the effect of
plotting distance vs.~distance in flux limited samples? In doing so, we
present further evidence that a strong {\em non-linear} correlation
among $L_{\rm R}$, $L_{\rm X}$ and $M$ indeed exists, which is {\it
not} affected by the range of distance and the heterogeneity of the
sample selection criteria.

\section{Fundamental plane vs.~distance driven artifact}
\begin{figure}
\psfig{figure=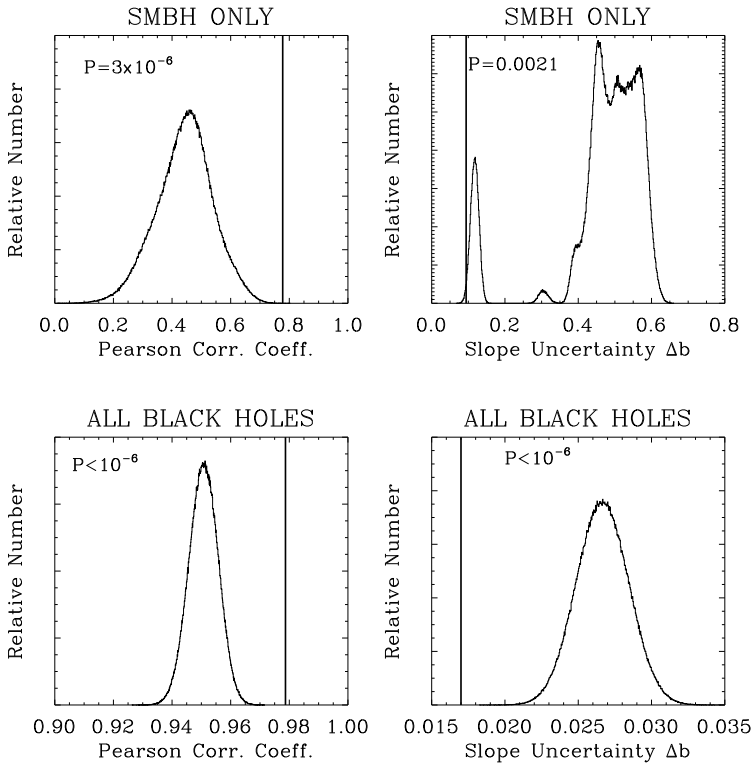,width=0.8\textwidth}
\caption{Results of the Monte Carlo simulation of scrambled radio
fluxes. {\em Upper panels}: extragalactic supermassive black holes
only; {\em lower panels}: entire sample of detected sources, including
XRBs; {\em left hand panels}: distributions of the Pearson's
correlation coefficients for randomized fluxes (curve), compared to
correlation coefficient of the original dataset (vertical line); {\em right
hand panels}: distributions of the uncertainties in the regression
slope for the randomized fluxes (curve), compared to the value for the
original data (vertical line).  Also shown are the Monte Carlo likelihood
values, $P$, for the observed values as random chance realizations of
the randomized sample (upper left corners).  All plots show clearly
that the randomized sample is not as strongly correlated as the real
one.}
\label{histo_all}
\end{figure}
Besides the formal partial correlation analysis, other rather straight-forward
tests can be easily carried out to check to what extent 
distance bias in our sample may be responsible for inducing the observed
correlation. For example, one can randomize the the observed fluxes
in any one band, and compare the correlation strengths of the original
and the randomized (``scrambled'') data\footnote{This specific test
was proposed by Bregman (2005)}.  The reason for this is obvious: if
the observed correlation is just an artifact introduced by the range
of distances in a sample of otherwise uncorrelated luminosities, then
the randomized datasets (the fluxes of which are guaranteed to be
intrinsically non correlated) should show the same degree of
correlation as the real dataset from which the fundamental plane was
derived.  Below, we will present a thorough, comparative statistical
analysis of the original sample with the randomized ones.

\subsection{The scrambling test I: SMBH only}
\label{sec:scram1}
We will first consider the extragalactic supermassive black holes
(SMBH) in the sample\footnote{Unlike the original sample by MHD03, we
remove the only genuinely beamed source (3C 273) for consistency}.  If
we consider only the detections (79 objects) and exclude the upper and
lower 5\% in radio luminosity, the sample spans a 90\% range of $\log
F_{\rm R,max}-\log F_{\rm R,min} \simeq 3.6$ orders of magnitude in
radio luminosity and of $\log F_{\rm X,max}-\log F_{\rm X,min} \simeq
3.3$ in X-ray luminosity (see figure~\ref{fig:sample}). 
The range of distances spanned by the SMBH sample is also
significant. The 90\% range in the distances of the detected objects
is 85, so that the factor distance squared, that enters in the
luminosity has a range of about 7.2$\times 10^3$, which is of the same
order as the range in fluxes.  As argued by \cite{kembhavi:86}, a
comparable spread in distance should prevent a spurious
luminosity-luminosity correlation from dominating a strong, underlying
correlation signal.  However, it is clear that care has to be taken
when studying luminosity-luminosity correlations and that distances
effects should always be accounted for.

To test whether distance bias dominates the correlation
we take the radio fluxes of the detected sources
and randomize them by assigning radio fluxes to objects in the sample
via random permutations.
To construct this Monte Carlo test, we repeat this procedure $10^6$
times and calculate the Pearson correlation coefficient between
$L_{\rm R}$ and $0.6L_{\rm X}+0.78M$ for each of the randomized
datasets (using the code {\tt slopes}, developed by M.~Akritas \&
M.~Bershady {\tt http://astrostatistics.psu.edu/statcodes}).  The
upper left panel of figure \ref{histo_all} shows the distribution of
the correlation coefficients obtained from the randomized datasets.
For comparison, the correlation coefficient ($R\simeq 0.7775$) of the
{\it actual}, observed SMBH sample is marked by a vertical line.

The figure shows that, as expected, the range of distances in the
sample does induce at some level a spurious correlation, as the
distribution of R is peaked at positive values. However, if the
correlation seen in the real dataset were purely due to this spurious
effect, its Pearson correlation coefficient would lie within the
distribution of the scrambled data, which is clearly excluded by our
Monte Carlo simulation.  Out of a million realizations of the
randomized radio flux distribution, only 3 had a larger correlation
coefficient than the real data. Clearly, the real data are much more
strongly correlated than the scrambled data.

We also performed a linear regression on the scrambled data, with
slope $b$ and intercept $a$, using a "symmetric" fitting algorithm
(see MHD03, \S3.1)\footnote{In particular, we have used here both the
OLS bisector and the reduced major axis method as described in Isobe,
Feigelson, Akritas and Babu (1990) and in Feigelson \& Babu (1992),
and implemented in the code {\tt slopes}; figure \ref{histo_all} shows
only the results for the reduced major axis method, but the results are
consistent in the two cases.}.  The upper right hand panel of
figure~\ref{histo_all} shows the uncertainty in the derived value for
the slope $b$, which can itself be regarded as a measure of the
intrinsic scatter of the fitted data. Only in about 0.2\% of the
scrambled datasets was this uncertainty smaller than that obtained for
the real sample.  This confirms the statement made in MHD03 (derived
from partial correlation analysis), that the degree of correlation
among $L_{\rm R}$, $L_{\rm X}$ and $M$ cannot be dominated by the
effect of distances.

\subsection{The scrambling test II: SMBHs and XRBs}
\label{sec:scram2}
Next, we consider the entire sample of detected sources, including
XRBs, bringing the sample up to 117 points in total. It is obvious
that when the XRB in our own Galaxy are included the range of
distances spanned by the sample increases dramatically.  The 90\%
ranges in $\log F_{\rm R}$, $\log F_{\rm X}$ and $D^2$ are now,
respectively, 4, 5.7 and 4.6$\times 10^{10}$.

As for the SMBH sample discussed above, we performed a Monte Carlo
simulation by randomizing the radio fluxes of the entire sample
$10^{6}$ times.  The distribution of the resulting correlation
coefficients for the scrambled dataset (including XRB) is shown in the
lower right panel of figure \ref{histo_all}.

As expected, this distribution is now peaked at very high values of R,
demonstrating that indeed the large range in distances can induce a
spuriously strong correlation.  This effect is unavoidable when
comparing SMBH and XRB, and it is not going to improve with any volume
limited sample of extragalactic sources, as current
telescope sensitivities are still far from what would be required in
order to observe XRB down to low luminosities in nearby AGN hosts (see
below).

What is striking about the Monte Carlo results derived from the
combined sample is that the Pearson correlation coefficient of the
actual dataset (R=0.9786)
is even {\it more} inconsistent with the randomized data than in the
SMBH-only case.  Out of a million realizations of the randomized data
sets, {\em not even one} showed a stronger correlation than the real
data.  In other words, the probability that the correlation found by
MHD03 is entirely due to distance effects is less than $10^{-6}$.
This statement is confirmed by the distribution of the uncertainties
in the regression slope, shown in the lower right panel.

This is partly due to the fact that in the XRB sample, the radio and
X-ray luminosities are correlated quite tightly, over a range of
luminosities much larger than the range in distances out of which they
are observed (see e.g.~Gallo et al.~2003).  More importantly, the
X-ray fluxes of the XRBs are systematically enhanced compared to the
AGN X-ray fluxes, while the radio fluxes of both samples are
comparable.  In other words, the correlation is {\it non linear}
($L_{\rm R}\propto L_{\rm X}^{0.7}$) and the {\em slope} of the XRB
correlation is, within the errors, consistent with being {\em the
same} as that derived from the best fit of the SMBH only sample. It is
thus {\em a fortiori} consistent (within the uncertainties imposed by
the significant residual scatter) with the correlation that is derived
for the entire sample.

If the effect were purely distance driven, one would expect to find a
correlation slope much closer to linear (see
\S\ref{sec:nonlin}). The non-linearity between $L_{\rm R}$ and
$L_{\rm X}$ and the fact that the power-law index
 is the same for XRBs and SMBHs
produces a very strong signal in the correlation analysis, much
stronger than the spurious one induced by the distance effects (only
the latter can be recovered from a sample with scrambled radio
fluxes), at greater than the 99.9999\% level.

This simple test leaves little room for arguing that
the "fundamental plane" correlation between radio luminosity, X-ray
luminosity, and black hole mass does not exist and that instead is
induced entirely by distance bias. These results are consistent with the partial
correlation analysis by MHD03, where non-parametric tests
were used to handle censored data.

The fact that the correlation is stronger when XRB are included rather
than in the SMBH sample alone, {\it even after the effect of distances
is considered} can be hard to visualize when plotting the entire
fundamental plane. Such a difficulty amounts to that of distinguishing
two correlations, one with a Pearson correlation coefficient of
$R\approx 0.94$, another with $R\approx 0.98$, extending over more
than 12 orders of magnitude\footnote{If two samples of 117 data each
have two measured Pearson correlation coefficients of 0.94 and 0.98
respectively, then it is possible to show that the probability of the
former being intrinsically a better correlation than the latter, is of
the order of 10$^{-5}$, see Num.~Rec.~chapter 14}.  We believe that
the difficulty in visualizing this statistically significant
difference may induce some concern on the fundamental plane
correlation. As we have shown, however, an accurate
statistical analysis can easily reveal this difference.
This visualization difficulty
also explains why the few upper limits in the MHD03 sample, when
plotted against the entire fundamental plane, will follow the same
correlation. A better test in this case would be to {\it quantify} the
degree of such a correlation for the censored data in the sample. 
The scrambling tests suggest
that they will indeed be correlated, but not as strongly as the real
dataset. There are, however, too few upper limits in the SMBH sample
of MHD03 to allow a meaningful statistical test.

\subsection{On the effects of flux limits}
Another way to test whether the fundamental plane is a pure distance
artifact
is to explore the flux selection effects using a Monte Carlo
simulation under the null-hypotheses that 
the radio, X-ray luminosities and black hole masses are not
correlated 
and assume we have a purely flux limited sample (however, see
section~\ref{sec:sample} below for a more accurate discussion of the
actual sample selection). 

\begin{table*}
\label{estimators}

\caption{Effects of the observing flux limits on uncorrelated data. The
first column is the radio flux limit in mJy, the second the X-ray flux
limit in units of $10^{-13}$ erg s$^{-1}$ cm$^2$. Third, fourth and
fifth column show the fitting parameters of the artificial sample with
a linear relationship $\log L_{\rm R}= \xi_{rm RX} \log L_{\rm X} +
\xi_{\rm RM} \log M + b$ (see Eq. 1). $\sigma_{\perp}$ is the scatter perpendicular to
the fitted plane.  
The partial correlation coefficient, $R_{RX,D}$ measures the degree of
intrinsic correlation of radio and X-ray luminosity once the effects
of distance are taken into account.  \label{tabMCTest}} 
\vspace{0.3cm}
\begin{tabular}{lcccccc}
\hline \hline
 $f_r$ & $f_x$ &  $\xi_{\rm RX}$ & $\xi_{\rm RM}$ & $b$ & $\sigma_{\perp}$ & $R_{RX,D}$ \\
\hline
\multicolumn{5}{l}{Luminosity function: $\alpha_r = 0.78$ and $\alpha_x = 0.85$}\\
\hline
0.1 & 0.1 & $1.09 \pm 0.07$ & $ -0.05 \pm 0.09$ & $-6.7\pm 2.2$ & $0.54\pm 0.04$  & $0.04 \pm 0.09$ \\
0.5 & 1 & $1.01 \pm 0.05$ & $ 0.0 \pm 0.06$ & $-3.9 \pm 1.6$ & $0.53 \pm 0.05$ & $0.05 \pm 0.08$\\
0.5 & 10& $1.04 \pm 0.09$ & $ -0.05 \pm 0.10$ & $-5.9 \pm 2.8$ & $0.53 \pm 0.05$ & $0.05 \pm 0.09$ \\
5 & 10& $1.03 \pm 0.05$ & $ -0.02 \pm 0.06$ & $-4.5 \pm 1.6$ & $0.51 \pm 0.05$ & $0.04 \pm 0.08$ \\
\hline
\multicolumn{7}{l}{Luminosity function: $\alpha_r = 0.55$ and $\alpha_x = 0.65$}\\
\hline
0.1 & 0.1 & $1.12 \pm 0.11$ & $ -0.09 \pm 0.13$ & $-7.3 \pm 3.7$ & $0.73 \pm 0.06$ & $0.01 \pm 0.08$ \\
\hline
\multicolumn{7}{l}{AGN Luminosity function: $\alpha_r = 1.5$ and $\alpha_x = 1.5$}\\
\hline
0.1 & 0.1 & $1.14 \pm 0.05$ & $ -0.06 \pm 0.04$ & $-8.5 \pm 1.5$ & $0.34 \pm 0.02$ & $0.09 \pm 0.07$ \\
\hline
\multicolumn{7}{l}{AGN Luminosity function: $\alpha_r = 0.78$ and $\alpha_x = 0.85$ XRBs correlated}\\
\hline
0.5 & 1 & $1.06 \pm 0.06$ & $ 0.32 \pm 0.04$ & $-8.7 \pm 2.1$ & $0.50 \pm 0.04$ & $0.26 \pm 0.09$ \\
5 & 0.1 & $1.22 \pm 0.09$ & $ 0.51 \pm 0.05$ & $-14.5 \pm 3.3$ & $0.54 \pm 0.05$ & $0.09 \pm 0.08$ \\
0.1 & 10 & $1.0 \pm 0.05$ & $ 0.12 \pm 0.04$ & $-6.6 \pm 1.8$ & $0.46 \pm 0.05$ & $0.35 \pm 0.08$  \\
5 & 10 & $1.1 \pm 0.07$ & $ 0.30 \pm 0.05$ & $-10.1 \pm 2.4$ & $0.48 \pm 0.04$ & $0.02 \pm 0.09$ \\
\hline \noalign{\smallskip}
\end{tabular}
\end{table*}

Radio and X-ray luminosity functions of AGN evolve strongly with redshift 
(see e.g., Hasinger, Miyaji and Schmidt 2005 and references therein; 
Willott et al 2001), however, as the original samples of MHD03 and FKM04
were almost exclusively made of local sources, the inclusion of these 
effects is beyond the scope of this simple test.
The observed luminosity functions are more complex than a simple power
law, 
being usually described by a flat power law (index 0.5-0.9) at lower 
luminosities and steeper one towards higher luminosities 
(for the 2-10 keV hard X-rays luminosity function 
the transition takes place around $10^{44}$ erg s$^{-1}$, see 
Ueda et al. 2003). However, as the majority of our sources are nearby 
galaxies with low luminosities, we can simplify our model and describe 
the luminosity functions with a simple power law.
 
Thus, we assume that the density of X-ray emitting AGN, $\Phi_X$, 
can be written as $\Phi_{\rm X}(L_{\rm X}) \propto L_{\rm
X}^{-\alpha_x}$. 
Similarly, the radio luminosity function is chosen to be $\Phi_{\rm
R}(L_{\rm R}) \propto L_{\rm R}^{-\alpha_r}$.
As a reference model, we fixed $\alpha_x = 0.85$ (Ueda et
al. 2003) and $\alpha_r \approx 0.78$ (Nagar et al 2005), and 
we will discuss below how our results depend on the exact slopes of 
the luminosity functions. 
We assume that the objects have a constant space density. 
For simplicity, we first assume that the luminosity functions 
for XRBs is the same as for AGN, though
this will be corrected in a second step.

We then construct an artificial sample containing 50 XRBs and 51 
nearby low luminosity AGN and 48 distant AGN and restrict the 
distances to the range observed: for XRBs 2 - 10 kpc, 
for LLAGN 3-50 Mpc, and for AGN 50-1000 Mpc. 
The average mass of an XRB is set to 10 $M_\odot$, with the masses
normally 
distributed in log-space with a dispersion of 0.3 dex. 
For AGN, we assume an average mass of $10^8 M_\odot$ 
and a dispersion of 1 dex. 
We always assume a distance measurement error of 10\% and an error in 
the mass estimate for the supermassive black holes of 0.35 dex.

This artificial sample can now be observed with given flux limits. 
As $\alpha_x,\alpha_r >0$, 
brighter objects are more likely to be detected 
at larger distances as the available volume is larger;
obviously, fainter sources cannot be detected out to these distances
due to the flux limits of the sample.
 
We have then performed a correlation analysis on these simulated
samples, varying both the flux limits and the slopes of the luminosity
functions.
The results are shown in table~\ref{tabMCTest}. 
All  samples show strong correlations 
(Pearson corr. coefficient $\approx 0.95$), indeed consistent with the
results of the scrambled samples discussed in section~\ref{sec:scram2}. 
However, not a single setup is able to yield the parameters 
similar to those of the fundamental plane, neither in term of
correlation strength, nor in terms of linear regression slopes. 
In particular, as the correlation is only created by the flux limits, 
we find $\xi_{\rm RX} \approx 1$, as it is expected given that 
we are simply plotting distance against distance \citep{feigelson:83}. Also as expected, 
the flux limits have no effect on the mass term, 
and one therefore finds no mass dependence of the radio luminosity
$\xi_{\rm RM} \approx 0$. 

When we perform on our artificial samples a partial correlation
analysis, i.e. we study the strength of the observed correlation {\it
taking distance into account}, as it was done in MHD03. We found 
that the partial correlation coefficient, $R_{RX,D}$ is always 
compatible with being zero, while the observed fundamental plane 
has a partial correlation coefficient 
of about 0.6. 
If one decreases the power law indexes for the luminosity functions, 
the perpendicular scatter $\sigma_{\perp}$ increases, but the slope 
of  the spurious correlation remains the same. 
We can therefore safely 
reject the null-hypotheses.

As discussed above, 
for XRBs it has been shown that in all low/hard state objects 
the radio and X-rays are correlated (Gallo et al. 2003). 
Thus, the question arises of whether the fundamental plane relation
can be a spurious effect generated by combining the genuinely
correlated sample of XRB with a flux limited, uncorrelated sample of
nearby SMBH.
 
The results are also shown in table~\ref{tabMCTest}. Again, 
the parameters are not compatible with those of the fundamental
plane. 
The partial correlation coefficient is now bigger than zero, 
as the XRBs are indeed correlated (but not the AGN), but it is still
much lower than for the real MHD03 sample. 
Thus, as the exact shape of the luminosity functions does not 
seem to change our results, this result further supports  the idea that
 the radio and X-ray luminosities of accreting black holes, 
as well as their masses are genuinely correlated.

\begin{figure}
\begin{tabular}{cc}
\psfig{figure=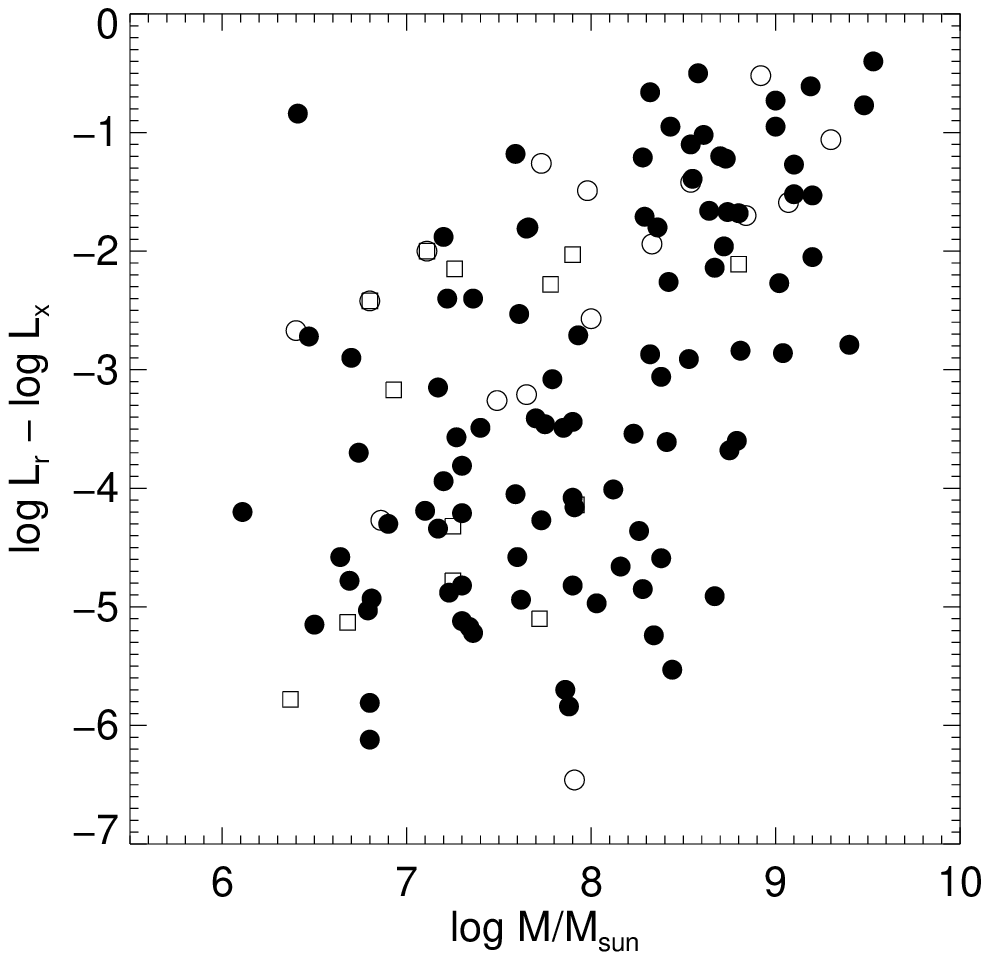,width=0.45\textwidth}&
\psfig{figure=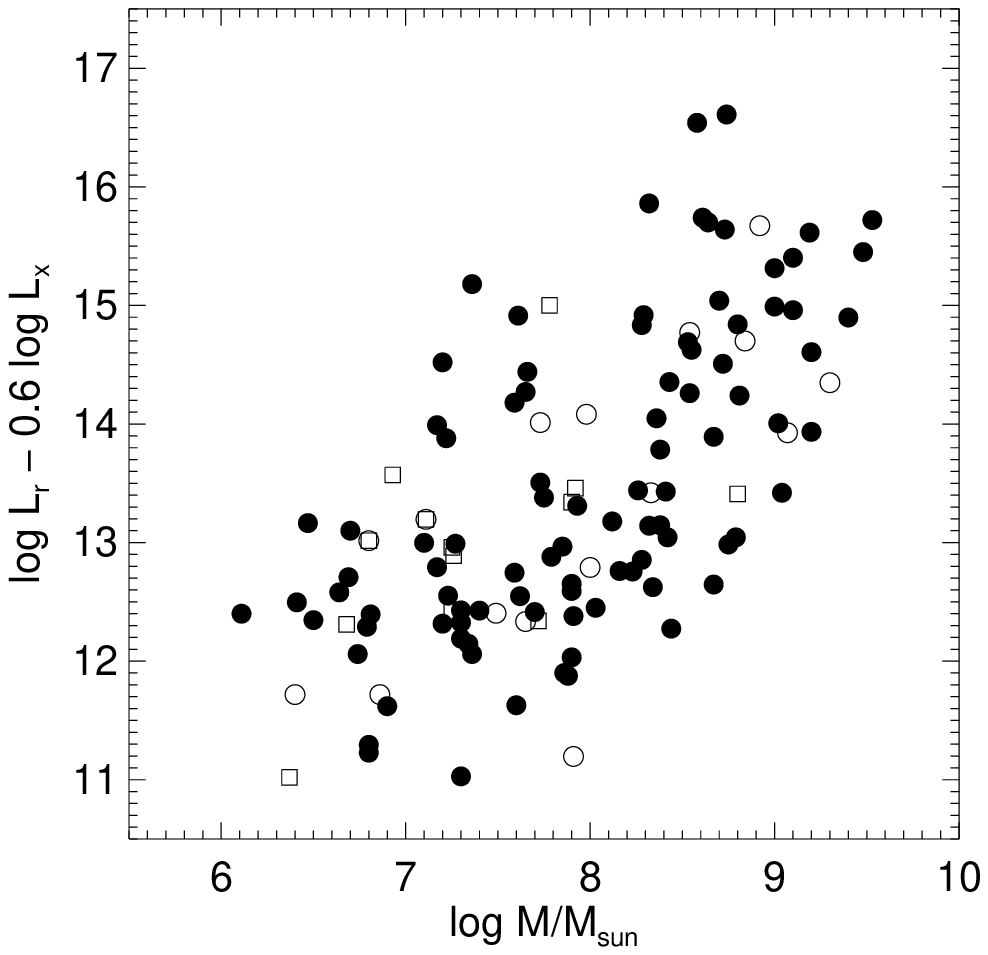,width=0.45\textwidth}
\end{tabular}
\caption{The left panel shows the logarithm of the ratio of radio to
  X-ray luminosity, $log L_{\rm R} - log L_{\rm X}$ vs. the logarithm
  of black hole mass for the SMBH in the sample. The right panel shows
  instead the ratio $L_{\rm R}/L_{\rm X}^{0.6}$. The latter shows
  clearly a stronger correlation with black hole mass than the former.
  Note that this plot removes the distance bias up to the level that
  black hole mass is only very slightly correlated with distance
  within the extragalactic SMBH sample (this can be seen from
  Fig.~\ref{fig:massdistance}). Open symbols are upper limits.}
\label{fig:lumlum}
\end{figure}

\subsection{Distance independent plots}
Obviously, it is possible to remove the distance effect entirely from
the analysis of any sample. 
If one were to expect a linear relation between $L_{\rm R}$
and $L_{\rm X}$, and some combined dependence of both on $M$, one
could, for example, plot $L_{\rm R}/L_{\rm X}$ vs.~$M$, in which case
the common distance dependence of $L_{\rm R}$ and $L_{\rm X}$ is
removed.

However, as was explained at length in MHD03, and as should be
apparent from the well known radio/X-ray relation in XRBs, one should
{\em not} expect a priori that the relation between the two is linear.
Rather, it is reasonable to expect that, to lowest order, the two will
follow a non-linear relation of the form $L_{\rm R} \propto L_{\rm
X}^{\xi_{\rm RX}}$ (though the exact power-law index $\xi_{\rm RX}$ of
this non-linearity depends on model assumptions).

\begin{figure}
\psfig{figure=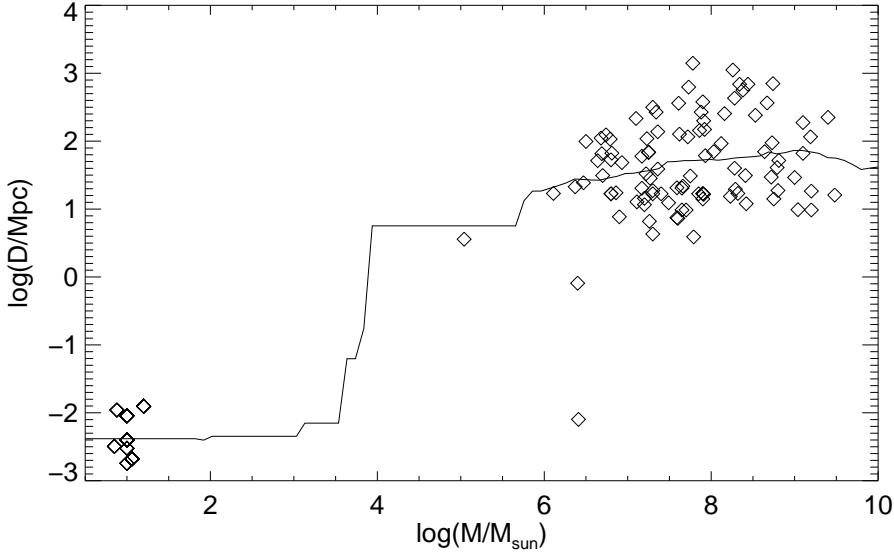}
\caption{Distance vs. black hole mass for the objects in the MHD03
  sample. the solid line is the sliding mean. This shows that the AGN
  sample is homogeneous in distance with mass and therefore any
  $L_{\rm R}/L_{\rm X}^{\xi_{\rm RX}}$ vs. M relation in the AGN
  sample cannot be driven by distance.}
\label{fig:massdistance}
\end{figure}

This suggests that a better variable to plot would be $L_{\rm
R}/L_{\rm R}^{\xi_{\rm RX}}$ vs.~$M$.  Using the best fit value of
$\xi_{\rm RX} = 0.6$ from MHD03, this is shown in
Fig.~\ref{fig:lumlum}, where it is compared to the same plot if a linear relation
between $L_{\rm R}$ and $L_{\rm X}$ is assumed.  Clearly, the
non-linear plot is significantly more correlated than the linear plot.
Note that this plot removes the distance bias up to the level that
black hole mass is only very slightly correlated with distance within
the extragalactic SMBH sample (this can be seen from
Fig.~\ref{fig:massdistance}).  This statement can be quantified: The
correlation coefficient for the two variables $L_{\rm R}/L_{\rm
X}^{\xi_{\rm RX}}$ and $M$ has a maximum of 0.65 at $\xi_{\rm RX} \sim
0.5$, compared to the value of $R=0.4$ reached at $\xi_{\rm RX}=1$
(note that this correlation does not use a symmetric method, thus
resulting in a different value than the $\xi_{\rm RX} \sim 0.6$ found
in the regression analysis of MHD03).  This difference is significant
to the 99.99\% level.

\begin{figure}
\begin{tabular}{cc}
\psfig{figure=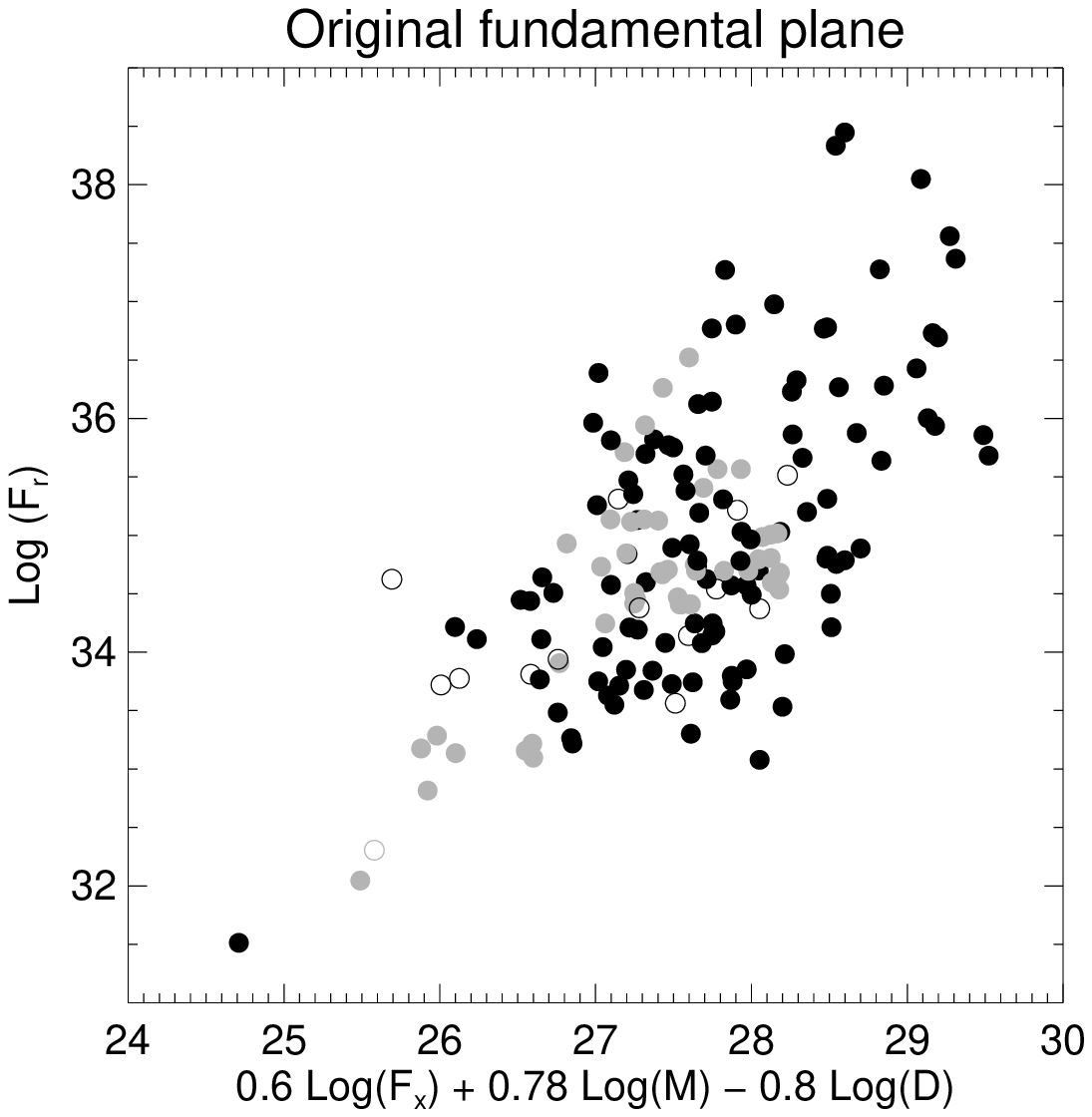,width=0.5\textwidth}&
\psfig{figure=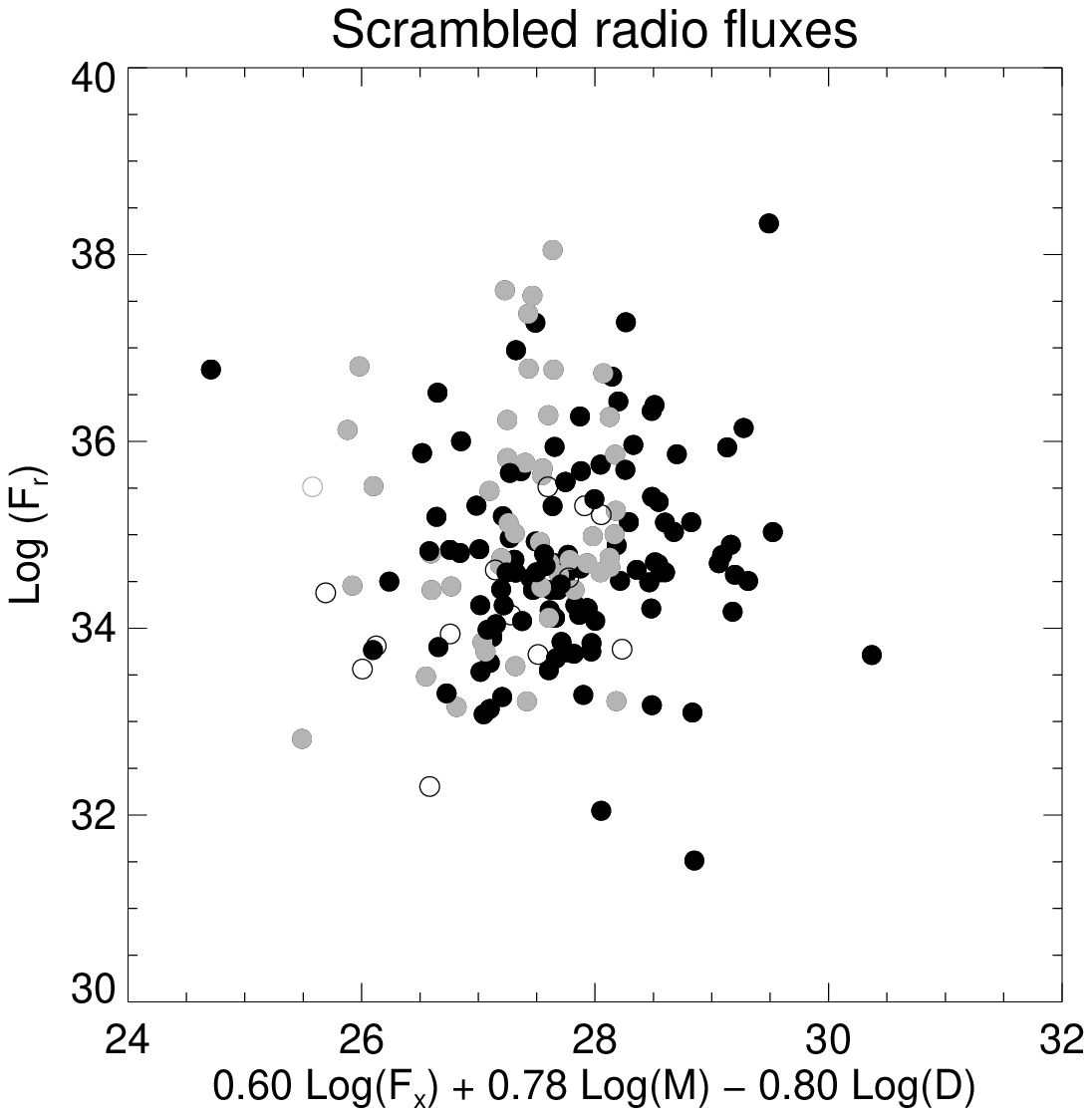,width=0.5\textwidth}\\
\psfig{figure=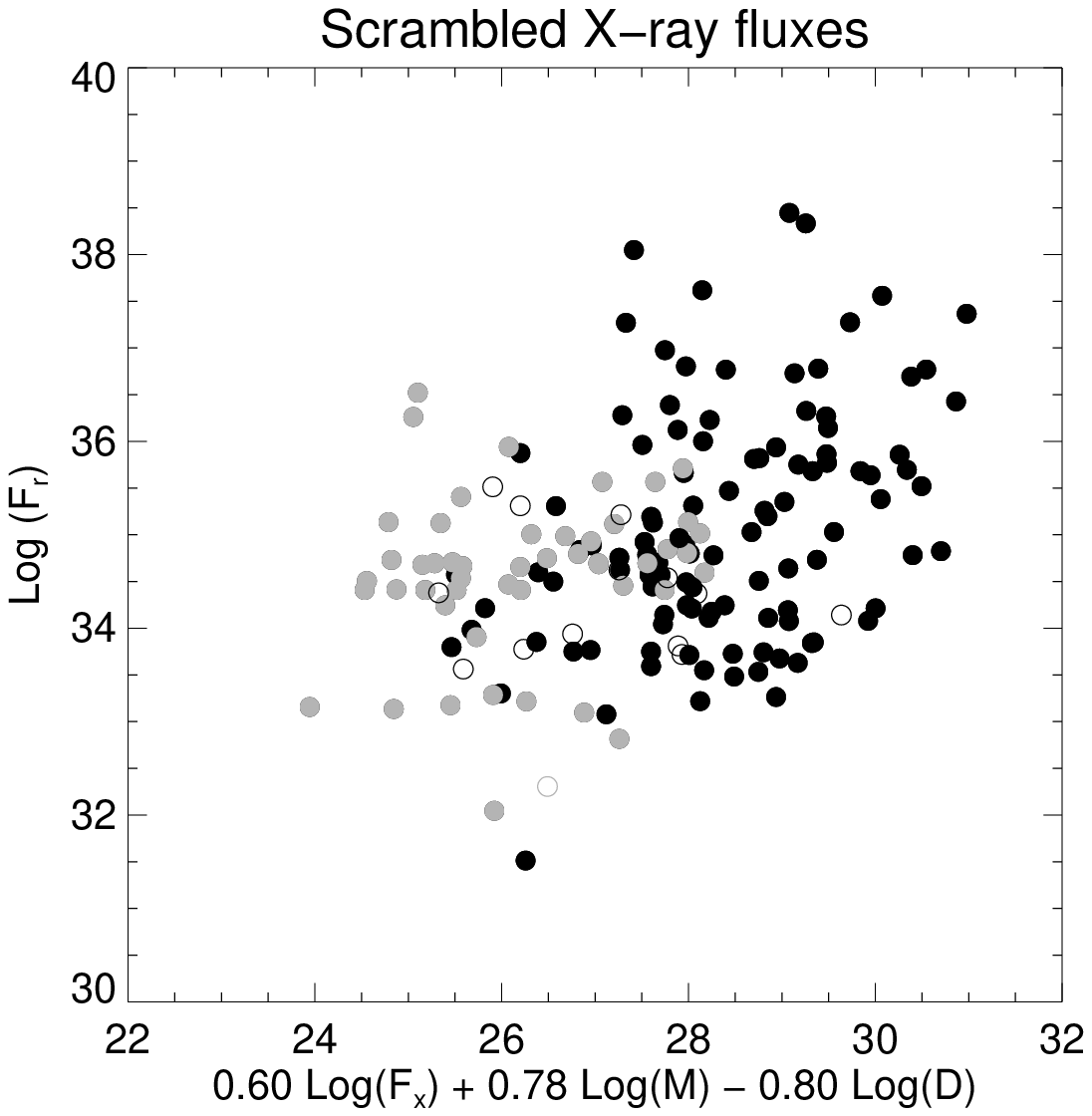,width=0.5\textwidth}&
\psfig{figure=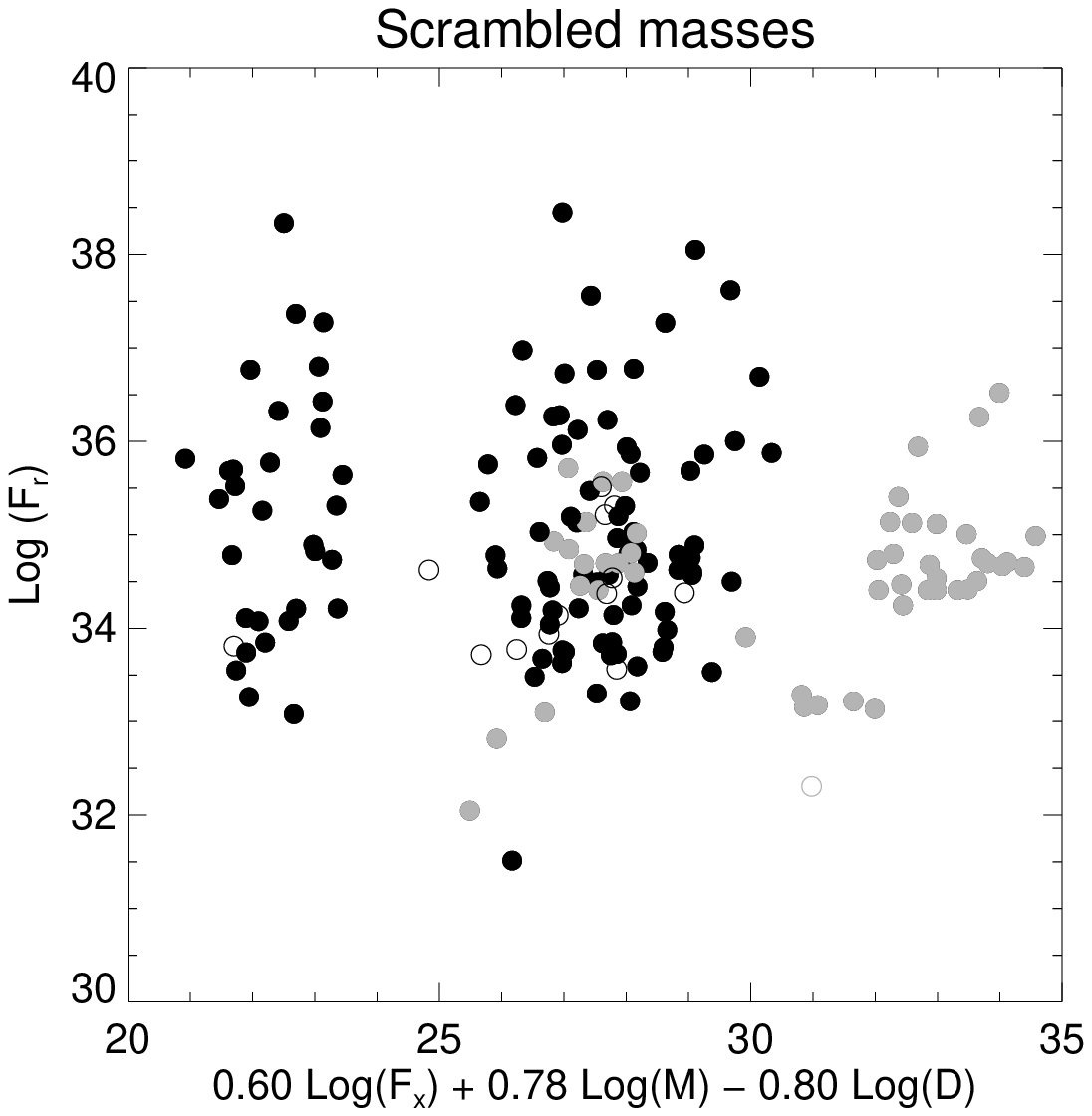,width=0.5\textwidth}\\
\end{tabular}
\caption{The upper left panel shows the fundamental plane relation in
  a flux-flux-distance, rather than luminosity-luminosity plot (fluxes
  are calculated measuring distances in Mpc). The
  other three show the same dataset in which either radio flux, or
  X-ray flux or mass has been randomized. Filled black symbols are for
  SMBH, filled grey ones for XRB and open symbols for upper limits.}
\label{fig:fp_d_all}
\end{figure}

Yet another related, visually clear, illustration of the fact that the
fundamental plane correlation is much stronger than any distance
induced bias can be shown by plotting the data in the flux-flux-mass
space.  Figure~\ref{fig:fp_d_all} shows in the upper left panel the
data viewed across the fundamental plane relationship expressed in
fluxes and with the distance as a fourth variable. The correlation
found in MHD03, expressed this way, reads:
\begin{equation}
Log F_{\rm R} = 0.6 Log F_{\rm X} + 0.78 Log M - 0.8 D + 7.33
\end{equation}

The other three panels of Fig.~\ref{fig:fp_d_all} show the data points
after a randomization of radio fluxes (upper right panel), of X-ray
fluxes (lower left panel) and of black hole mass (lower right panel).
A visual inspection is sufficient to show that the correlation in the
original data is much stronger than the residual correlation in the
lower left panel (scrambled X-ray fluxes - note that a residual
correlation should be expected in this case, as the radio luminosity
should be related to black hole mass even for a random set of X-ray
luminosities) and that no correlation is present in the other two panels.
By construction, this correlation cannot be a spurious distance
effect.

\subsection{On sample selection}
\label{sec:sample}
Clearly, the plots in Fig.~\ref{fig:fp_d_all}, as well as the
confidence in the regression slopes, could be improved by a more
carefully crafted, more complete sample than what we have currently
available. We shall briefly address the question of whether a volume
limited sample would, in fact, be the best way to treat this problem,
as advocated, for example, in Bregman (2005).

In what follows, it is important to keep in mind that the
original sample studied in MHD03 was {\it neither} a flux limited
sample, {\it nor} a combination of flux limited samples, but rather a
combination of flux and volume limited samples, observed in both X-ray
and radio bands with different sensitivities (see
figure~\ref{fig:sample}). For example, MHD03 considered all known
Low-Luminosity AGN within 19 Mpc observed by Nagar et al. (2002) with
the VLA.  Upper limits were recorded as far as possible, whenever the
information regarding a source with reasonably well measured/estimated
black hole mass was available from radio or X-ray surveys, but no
effort was made to account for the incompleteness derived from the
requirement of a source having a measured black hole mass itself.  The
heterogeneity of the resulting sample may well introduce biases which
are hard to account for in a luminosity-luminosity correlation;
however, it is also a safeguard against systematic effects that might
arise from any one technique of estimating black hole masses.

Furthermore, the two populations
have vastly different distances, masses, and luminosities.  Clearly,
these distinct regions of parameter space are largely responsible for
stretching out the original plot of the fundamental plane over fifteen
orders of magnitude on each axis.  The question then arises whether a
volume limited sample could address some of the concerns about
spurious distance effects discussed above
(after all, even the randomized data show a correlation coefficient of
0.94).
Before addressing this question, however, it is important to note that it is
not at all unreasonable to compare X-ray binaries and AGNs in the same
flux range, and that a volume limited sample including both XRBs and
SMBHs would, in fact, not make much sense.  Physical intuition
suggests that, when comparing black holes of vastly different masses,
one should restrict the analysis to a similar range in dimensionless
accretion rate, $\dot{m}\equiv \dot{M}/M$.  By coincidence, the
roughly seven orders of magnitude difference in $M$ between XRBs and
SMBHs are almost exactly canceled out by the roughly 3.5 orders of
magnitude larger distance to the SMBH sample, making the flux ranges
spanned by XRBs at least comparable.  As it turns out, comparing the
volume limited XRB sample with flux limited AGN sample puts both
classes in roughly the same range of $\dot{m}$ (individual sources
like GX 339-4 and Sgr A* representing a small percentage of outliers).

In a volume limited sample that includes both AGNs and XRBs, one would
be forced to compare objects at vastly different accretion rates,
which would not be very meaningful from a physical point of view.  In
this sense, one could also argue that the distance bias that is
invariably introduced when correlating XRBs and AGNs is in reality an
accretion rate bias, which is warranted on physical grounds.

Furthermore, due to the cosmological evolution of the accreting black
holes population, a volume limited sample would be strongly dominated
by quiescent sources for AGNs.  For fitting regression slopes, a
sample crafted to have roughly equal density of points throughout the
parameter space would presumably
be much better suited for determining the regression slopes.  While the MHD03
sample is certainly far from reaching that goal, it is another
argument against a broad brush call for volume limited samples.

\section{The slope of the fundamental plane}
\label{sec:nonlin}
Fitting a regression through the data requires the {\em assumption}
that one single underlying relation drives the data.  Within that
context, the regression will produce the correct slopes no matter what
the sample is. The same is true for including XRBs: although 
they may have comparable slope to the AGNs and although the AGNs lie on
the extrapolation of the XRB slope with the mass correction, the
statement
that these facts are truly
an expression of the same accretion physics at work must be posited as
an ansatz (see MHD03).

The fact that radio/X-ray 
correlations can be found in samples of XRBs and AGN, 
either in luminosity-luminosity, or flux-flux
(with slaved distance) space that are consistent with each other
within the uncertainties then supports the ansatz, and the correctness
of the idea of jointly fitting {\em one} correlation.  Within those limits,
the slopes we derived are an accurate representation of the putative
relation. This is in fact the customary and correct way to
proceed. First one should test that the available data are indeed
correlated, taking all possible biases (as those induced, for example,
by distance, sample selection, etc.) into account. If, and only if,
any such correlation is found to be statistically significant, then a
linear (possible multivariate) regression fit to the data can be
looked for.

Within the present context, the clear non-linearity of the correlation
between radio and X-ray luminosity for XRB and the apparent
non-linearity of the correlation for the SMBH-only sample (with the slope
consistent with being the same in the two separate samples), not only
reinforce strongly the validity of our approach, but also suggests that
only by working in the luminosity-luminosity space can one recover the
intrinsic properties of the objects under scrutiny 
\citep{feigelson:83,kembhavi:86}.

\subsection{Using simultaneous radio/X-ray observations}
\label{sec:m81}

\begin{figure}
\psfig{figure=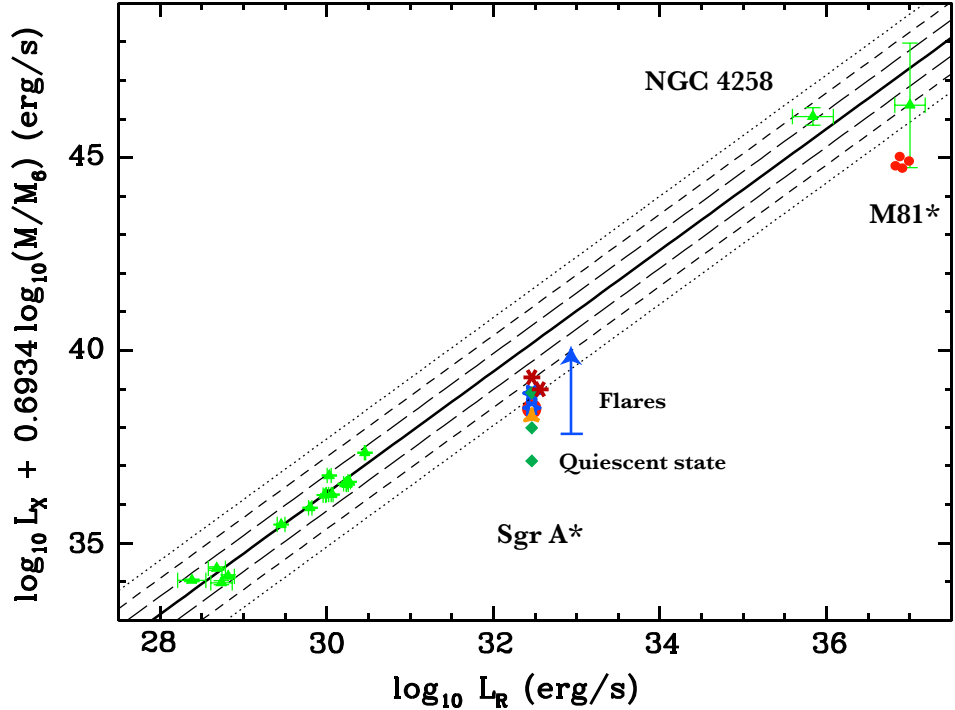,width=0.9\textwidth}
\caption{The fundamental plane correlation for the three well-measured
sources bracketing Sgr A*: the X-ray binary GX 339-4, and the LLAGN
NGC 4258 and M81. For this latter source the red points represent last year's
simultaneous observations (Markoff et al., in prep.), while the green point with large error
bars is the average and rms
variation from all prior
non-simultaneous observations. The solid line indicates the best fit correlation
from Monte Carlo simulations (see Markoff 2005 for details), with
contours in the average scatter $\langle \sigma \rangle$ from the
correlation represented as increasingly finer dashed lines.}
\label{fig:markoff}
\end{figure}

As a final note, it is useful to point out one additional factor which
should be considered when searching for the true nature (i.e. slope) of
the fundamental plane.  In XRBs, discovering and properly measuring
the non-linear radio/X-ray correlation depended on the existence of
good quality, quasi-simultaneous radio and X-ray band observations of
a single source.  The luminosity changes during outburst cycles which
trace out this correlation occur on timescales of days to weeks. Radio
and X-ray flux measurements separated in time by more than this 
would result in an
altogether different correlation reflecting the lag in observation
time.  Our techniques of testing whether AGN follow a similar correlation
as XRBs by using samples are viable only with the inherent assumption
that non-simultaneous radio/X-ray observations are comparable for
these sources.  In general, this should be true because AGN are
expected to vary on longer timescales than XRBs, by a
factor that scales roughly linearly with their mass ratio.  In other
words, if a single AGN observation is equivalent 
to a single data point on the XRB radio/X-ray correlation, then we assume
that we would in fact see the same type of
correlation if we could study an AGN for millions of years.

Nearby low-luminosity AGN (LLAGN), however, often have smaller central
masses ($\sim 10^{6}-10^{7} M_{\odot}$
) and can show variations in both radio and
X-ray fluxes of tens of percent over month-long timescales.  As an
alternative test of the reality of the fundamental plane, one can study
the fit to the plane and its scatter using just a few sources with
very well measured mass and distance, so that the scatter is in fact
dominated by intrinsic variability.  An initial test was performed by
Markoff (2005), using data from the best XRB displaying the
radio/X-ray correlation in its hard state, GX 339-4, as well as our
Galactic, underluminous SMBH Sgr A*, and two nearby LLAGN, M81 and NGC
4258.  All of these sources have well-determined physical mass and
distance, and are not highly beamed, allowing a detailed assessment of
the fundamental plane coefficients, as well as the contribution to its
scatter from intrinsic variability.  A linear regression fit was
performed on $10^4$ "samples" of data, simulated using a Monte Carlo
technique from decades of observations of the sources, and representing
all possible configurations of their respective fundamental plane
during different phases of variability.  The best fit plane is shown
in Fig.~\ref{fig:markoff},
 with contours in average scatter indicated, which was used
to estimate the relationship of Sgr A*'s flares to the fundamental
plane relation.  Interestingly, the resulting fundamental plane
coefficients are similar to those
derived by MHD03 and FKM04, 
although the mass-scaling factor $\xi_{\rm RM}$ is somewhat smaller.

In the past year, truly simultaneous observations of M81* have
been carried out (Markoff et al., in prep.) and four actual data
points like those for the XRBs can be added to the plane projection.
These have been placed on Fig.~\ref{fig:markoff} 
for comparison to the data point
representing the average and rms variation from all prior
non-simultaneous observations.  If these points had been used in a
single determination of the fundamental plane, along with the GX 339-4
data and the same average/rms variation from NGC 4258, linear
regression would give the relation:
\begin{equation}
Log L_{\rm R} = 0.586 Log L_{\rm X} + 0.656 Log M + 8.211 
\end{equation}
Again, the radio/X-ray correlation coefficient is very similar to the
results derived by MHD03.  Because this directly tackles the question
of simultaneity, it is in fact a very strong test of the predicted
mass scaling.  It is compelling that the correlation derived from such
well-measured sources is consistent with the relations derived
from both XRB and the AGN samples.

\section{Conclusions}
We have presented further statistical evidence that the fundamental
plane of black hole activity (i.e. the non-linear correlation between
radio core luminosity, X-ray luminosity and mass of accreting black
holes) is not an artifact due to an overlooked bias introduced by the
range of distances spanned by our samples.

Partial correlation analysis techniques capable of handling censored
data were already used in the original work of MHD03, following a
decades long tradition in the multiwavelength study of AGN and
QSOs. Here we have extended this analysis performing Monte Carlo
simulations of randomized radio fluxes and found results entirely consistent
with the partial correlation analysis. Additional Monte Carlo
simulations of combined flux limited samples of XRBs and AGN have also
demonstrated that the orientation of the fundamental plane (i.e. its
slope) cannot be reproduced by spurious distance driven effects.
Moreover, also distance-independent tests 
demonstrate that the fundamental plane
correlation is real and has a non-linear slope, which further suggests
that studying flux-flux relations only is {\it not} appropriate when
dealing with the data.

With respect to the traditional studies of correlations between
luminosities of AGN in different bands, the inclusion of a mass term
in the analysis imposes a very complex selection criterion on any
sample: mass can be estimated in a number of different ways, with
different degrees of uncertainties, and different degrees of
observational difficulty, so that it is almost impossible, at least
with the current data, to estimate the degree of incompleteness of any
black hole mass sample. With respect to this crucial aspect, we argue
that volume limited samples are not necessarily the best tools to
study and understand the physical origin of such a correlation, as the
cosmological evolution of the population of accreting black holes
introduces severe biases in the ($M$, $\dot m$) parameter space, which
also have to be taken into account. Simultaneous observations in the
Radio and X-ray bands of accreting SMBH at the low-mass end of their
distribution will be extremely useful in better determining the true
correlation coefficients of the fundamental plane and thus place
better constraints on its physical origin.

\end{document}